\documentclass[a4paper,11pt]{article}
\usepackage{pos}
\usepackage{wrapfig}
\usepackage{enumitem}

\title{Results and plans to apply interferometry to air shower observations at the Pierre Auger Observatory }
 \ShortTitle{Air shower interferometry at the Pierre Auger Observatory}

\author*[ab]{Pim van Dillen}

\affiliation[a]{IMAPP, Radboud University Nijmegen, Nijmegen, The Netherlands}
\affiliation[b]{NIKHEF, Science Park, Amsterdam, The Netherlands}

\onbehalf{for the Pierre Auger Collaboration$^c$}
\affiliation[c]{Observatorio Pierre Auger, Av.\ San Mart{\'\i}n Norte 304, 5613 Malarg\"ue, Argentina\\
Full author list: {\rm\url{https://www.auger.org/archive/authors_icrc_2025.html}}}

\emailAdd{spokespersons@auger.org}

\abstract{By analysing the radio emissions from air showers using interferometry, we can estimate their properties. In this contribution, we apply interferometry to reconstruct air-shower parameters based on measurements taken with the Auger Engineering Radio Array (AERA) at the Pierre Auger Observatory. This reconstruction method is achievable at AERA through precise clock synchronisation with a beacon and an accurate survey of the station locations. Interferometry has been applied to several thousand inclined air-shower observations for the first time, which allows for tests on the performance of air-shower geometry reconstruction, recovery of the radio signal from low-energy air showers, and methods to study the polarisation of the radio-emission mechanisms. Additionally, in this contribution, we will also provide an overview of efforts to enable interferometry for the recently installed radio detectors that are part of the AugerPrime upgrade.}

\ConferenceLogo{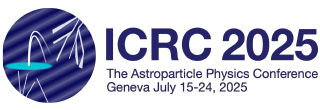}

\FullConference{39th International Cosmic Ray Conference (ICRC2025)\\
15 -- 24 July, 2025\\
Geneva, Switzerland}

\begin{document}
\maketitle
\section{Introduction}
When a high-energy cosmic ray enters the atmosphere, it creates a cascade of secondary particles called an extensive air shower. The electrically charged particles in these air showers emit radio waves, mainly caused by the deflection in Earth's magnetic field (for topical reviews, see~\cite{SCHRODER20171,ReviewHuege}). At the Pierre Auger Observatory~\cite{PAO}, the Auger Engineering Radio Array (AERA)~\cite{AERA} was built to detect this radio emission in the 30 - 80 MHz frequency range, allowing the reconstruction of various air shower parameters, such as the energy of the cosmic ray, the depth of the shower maximum (X$_\mathrm{max}$), and the arrival direction. With interferometry, the coherent properties of the radio signals are used to obtain a 3D mapping of the emission from the air shower. In~\cite{schoorlemmer-2021}, it has been shown that from simulations, it is possible to reconstruct the shower axis and X$_\mathrm{max}$. In~\cite{schoorlemmer-2023}, the interferometric technique was applied to a single event recorded by AERA, demonstrating the promising prospect of this method. In this contribution, we will show the results of the interferometric reconstruction on a large data set recorded by AERA of inclined air showers (zenith angles larger than 55$^\circ$).

\section{Interferometric air shower reconstruction}
\label{sec:algo}
\begin{wrapfigure}{r}{0.5\textwidth}
  \vspace{-10pt}
  \begin{center}
    \includegraphics[trim={0 0 0 0.1cm},clip,width=0.48\textwidth]{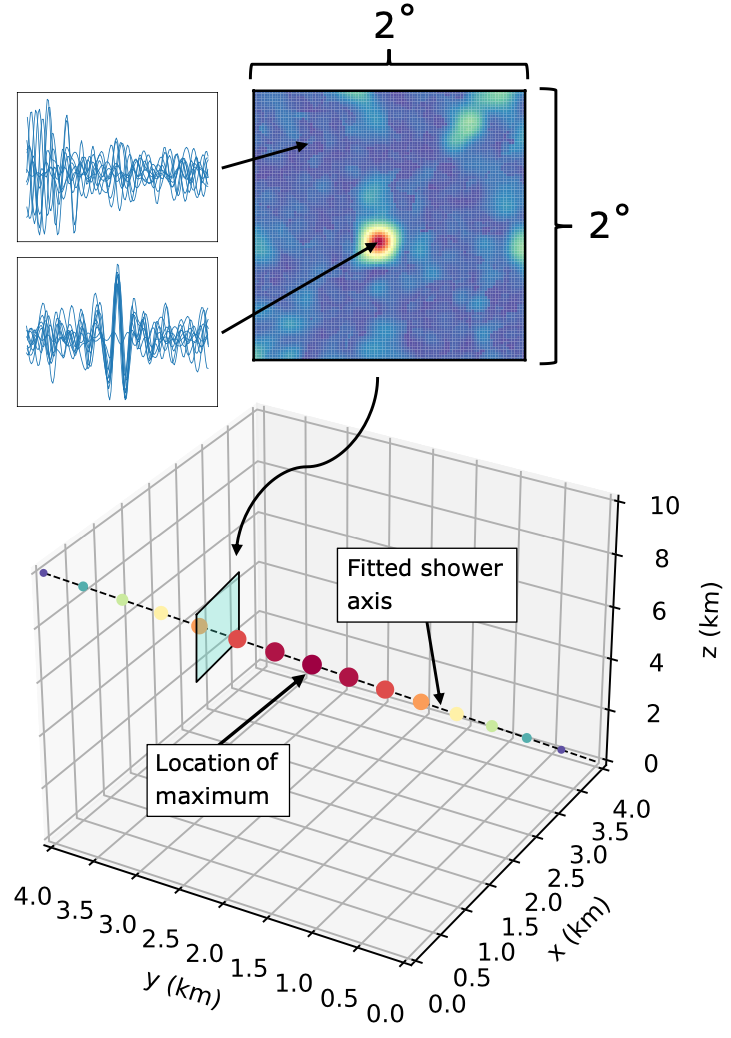}
  \end{center}
  \vspace{-10pt}
  \caption{Reconstruction using interferometry (see text for details)}
  \label{fig:schematic_reco}
\end{wrapfigure}
The algorithm used in this analysis generates a coherence mapping of the air shower by summing radio waveforms at a location in time and space that are corrected for the light travel time from that location to the receiving antennas.
In Figure \ref{fig:schematic_reco}, a schematic overview of the reconstruction of the air shower axis is given, and the details are described in the following. Using the Auger Offline software framework~\cite{ROffline,Offline}, we obtain the time-dependent electric field vector, which is projected on the polarization direction expected for the geomagnetic emission to obtain a waveform ($-\vec{v}\times\vec{B}$, with $\vec{v}$ the velocity of the particles in the air shower and $\vec{B}$ the Earth's magnetic field).\\ 
The shower axis is described by four parameters: the zenith angle $\theta$, the azimuthal angle $\phi$, and the intercept with a horizontal plane $x_0,y_0$, commonly called the shower core. The shower axis, as obtained from the air-shower reconstruction based on measurements from the surface detector, is used as a starting point for the algorithm. Perpendicular to this axis, we generate grids at regular intervals of slant depth, on which the intensity of the summed waveforms is evaluated. From each grid, we obtain the location and intensity of the maxima, which are used in a least-squares minimisation to obtain the (new) shower axis.  Along this reconstructed shower axis, we generate a large set of points and find the location of the maximum intensity. When expressed in slant depth, we refer to this location as X$_\text{RIT}$.

\subsection{Data set}
The data set consists of inclined (zenith angle > 55\,$^\circ$) air showers that are measured with both the 750\,m spaced array of the surface detector (SD-750) and the AERA. The events are triggered by the signals received by the surface detector, which initiates the read-out of waveforms of the antennas. 
For the interferometric method to perform well, the acquisition of radio measurement data must be synchronised with an accuracy of the order of a nanosecond, and the location of the antennas must be known within 30\,cm.  The clock synchronisation over the array is achieved for AERA by a beacon transmitter that continuously emits sinusoidal waves that are recorded with each waveform~\cite{Beacon_AERA}. In addition, the (relative) location of the antennas was surveyed using a differential GPS with an accuracy of $\sim$10\,cm.  

\subsection{Antenna Selection}
For the standard air-shower reconstruction of radio observations with AERA antennas, there is a signal threshold to identify whether a clear air-shower signal is present in the waveforms. However, with the interferometric approach, it is possible to use antennas in the reconstruction where the air-shower signal is not identifiable on a single waveform, as presented in~\cite{schoorlemmer-2021}. Antennas located far from the shower axis measure a negligible shower signal compared to the background noise, so to maximise the signal-to-noise ratio, an optimum was found for this analysis. At a reference depth of X = 750 g/cm$^2$, the Cherenkov angle was calculated and multiplied by 2.5, projecting an ellipse on the ground around the shower core, which was reconstructed by the information of the surface detector. All antennas within this area are used for the reconstruction. Additionally, to account for a mismatch between the real shower axis and the one reconstructed with the surface detector, also antennas with an identifiable signal close to 2.5 times the Cherenkov cone are included. The ellipse is enlarged by 0.5 times the Cherenkov angle, and if an antenna with identifiable pulse is located in the newly added area, the process continues. Otherwise, the antenna selection stops.\\
Within AERA, several station configurations~\cite{Antenna} were operated simultaneously. Since the interferometric reconstruction is (very) sensitive to the pulse shape, we decided to use only antennas that were equipped with so-called Butterfly antennas that were read out upon triggering from the surface detectors to reduce potential systematics from combining different systems. 

\subsection{Event Selection}
Not all events in the data set are suitable for reconstruction due to various reasons, e.g., thunderstorms, too few antennas, or noisy circumstances. Therefore, the total event set is reduced based on several selection criteria to improve the quality of the data set: 

\begin{enumerate}
    \item To ensure that the shower axis is identifiable above the incoherent background signal in the grids created in the reconstruction, a cut is placed on the \textit{mean slice intensity}. The \textit{mean slice intensity} is defined as the mean of the normalised intensities of the grid points without taking into account the 10 highest intensity points. This value is determined in each grid, and an event should have at least one \textit{mean slice intensity} below 0.2.
    \item To ensure that the air-shower axis fitted agrees with the points determined on the grids, we cut on the \textit{intensity residual}. The \textit{intensity residual} is defined as the normalised mean distance between the intensities of the points determined by the grids and the intensity on the reconstructed shower axis. This value should be lower than 0.1.
    \item The interferometric method works best if a large fraction of the radio footprint is sampled by antennas. To achieve this, it is required that the projection of the Cherenkov cone is contained in the area spanned by the antennas. 
    \item Thunderstorm conditions can affect the radio signal of air showers in unpredictable ways~\cite{lofar_thunderstorms}. At the Pierre Auger Observatory, the local electric field strength is measured, and a thunderstorm flag is given when the electric field exceeds a certain threshold, which indicates nearby thunderstorm cloud(s). To minimise this influence, events with thunderstorm flags are discarded.
\end{enumerate}
 
\subsection{Results on shower axis reconstruction and bias corrections}
\label{subsection: Results on geometry reconstruction and bias corrections}
\begin{figure}
    \centering
    \includegraphics[width=0.49\linewidth]{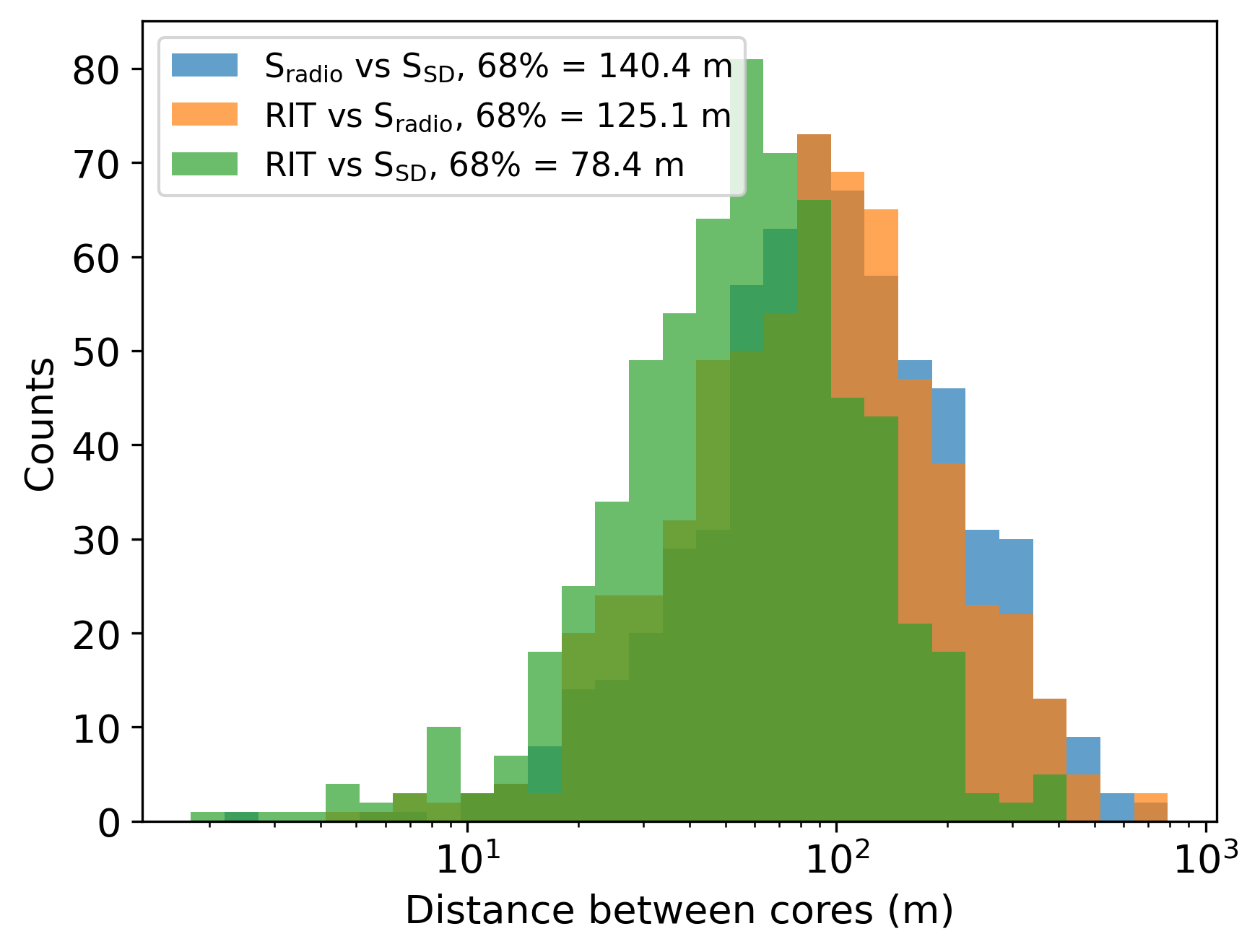}
    \includegraphics[width=0.49\linewidth]{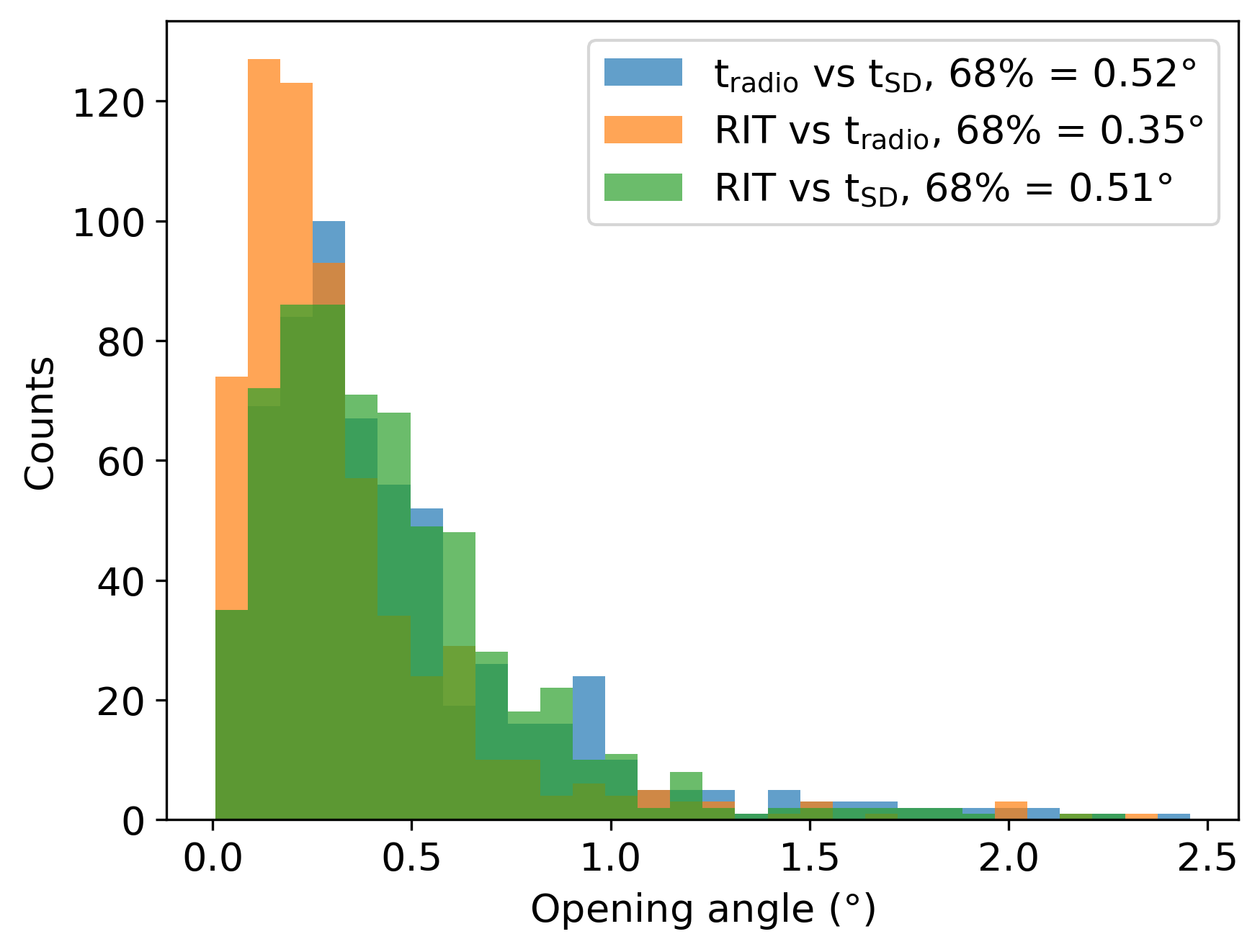}
    \caption{Comparing geometry reconstruction based on the surface detector (SD), the radio lateral distribution function (radio) and radio interferometry (RIT). \textit{S} and \textit{t} represent fit functions based on respectively, the signal strength and the timing of the signal for the corresponding detector. Left, the angular distance between the reconstructions of the directions of the axis. Right, the distance between the core reconstruction calculated in a plane perpendicular to the air shower axis.}
    \label{fig:reco_results}
\end{figure}

In Figure \ref{fig:reco_results}, the results for the reconstruction of the shower axis using interferometry (\emph{RIT}) are compared to the standard reconstruction for radio measurements with AERA (\emph{Radio}) and a custom reconstruction\footnote{adapted from the standard inclined event reconstruction for the SD-1500 described in \cite{SDReco}.} for inclined air showers with SD-750 detector (\emph{SD}). For the standard \emph{Radio} and \emph{SD} reconstructions, the direction of the shower axis is obtained by fitting an arrival time function, $t$, to the signal timing, and the core is obtained by fitting a lateral density function $S$ to the signal strength~\cite{SDReco}.
\\
By analysing the result of the interferometric reconstruction, several biases in the parameters were identified when compared to the SD reconstruction. These biases arise from the asymmetric distribution of the amplitude of the recorded radio signal with respect to the shower axis. The nature of this asymmetry is twofold; the radio signal is inherently asymmetric around the shower axis, due to interference of different emission mechanisms~\cite{2010APh....34..267D} and early-late effects in very inclined air showers~\cite{2020EPJC...80..643S}. However, in the case of observations with AERA, the asymmetry due to the recorded radio signals is dominated by the inhomogeneous and incomplete sampling of the footprint of the radio emission. To correct for these biases, we calculate for each recorded air shower the average position $(x_w,y_w)$ of the radio antennas in a plane perpendicular to the air shower axis weighted with the radio signal (also sometimes called the barycentre or the centre-of-mass) and define a correction based on the dependencies of $(\theta,\phi,x_0,y_0)$ on $(x_w,y_w)$. These corrections are obtained from simulated radio emission from inclined air showers (CoREAS/CORSIKA~\cite{CoREAS,corsika}) that are randomly distributed over AERA. The distributions in Figure \ref{fig:reco_results} have been corrected for these biases, resulting in reconstructions closer to the other methods.\\
In the distribution from the difference in the shower core reconstruction (left panel Figure \ref{fig:reco_results}), we observe that the radio interferometry method is closer to the \emph{SD} reconstructed variables than to the \emph{Radio} reconstructed variables, this might be (partially) due to an uncorrected bias in the core position for the \emph{Radio} reconstruction which is expected in inclined air showers but has not been corrected for~\cite{2020EPJC...80..643S}. For the direction reconstruction (right panel Figure \ref{fig:reco_results}), the \emph{Radio} reconstruction and \emph{RIT} are closest to each other and have a very similar difference with respect to the \emph{SD} reconstruction. We note that for readability, we did not display opening angles larger than 2.5$^\circ$, however about 15 events had significantly larger opening angles when comparing \emph{RIT-Radio} and \emph{SD-Radio}, an indication of a poor \emph{Radio} reconstruction. This can happen when a station is not well synchronised, or a noise pulse is accidentally used in the reconstruction; the radio interferometric reconstruction is (more) robust in these situations.

\section{Polarization}
\begin{figure}
    \centering
    \includegraphics[width=0.49\linewidth]{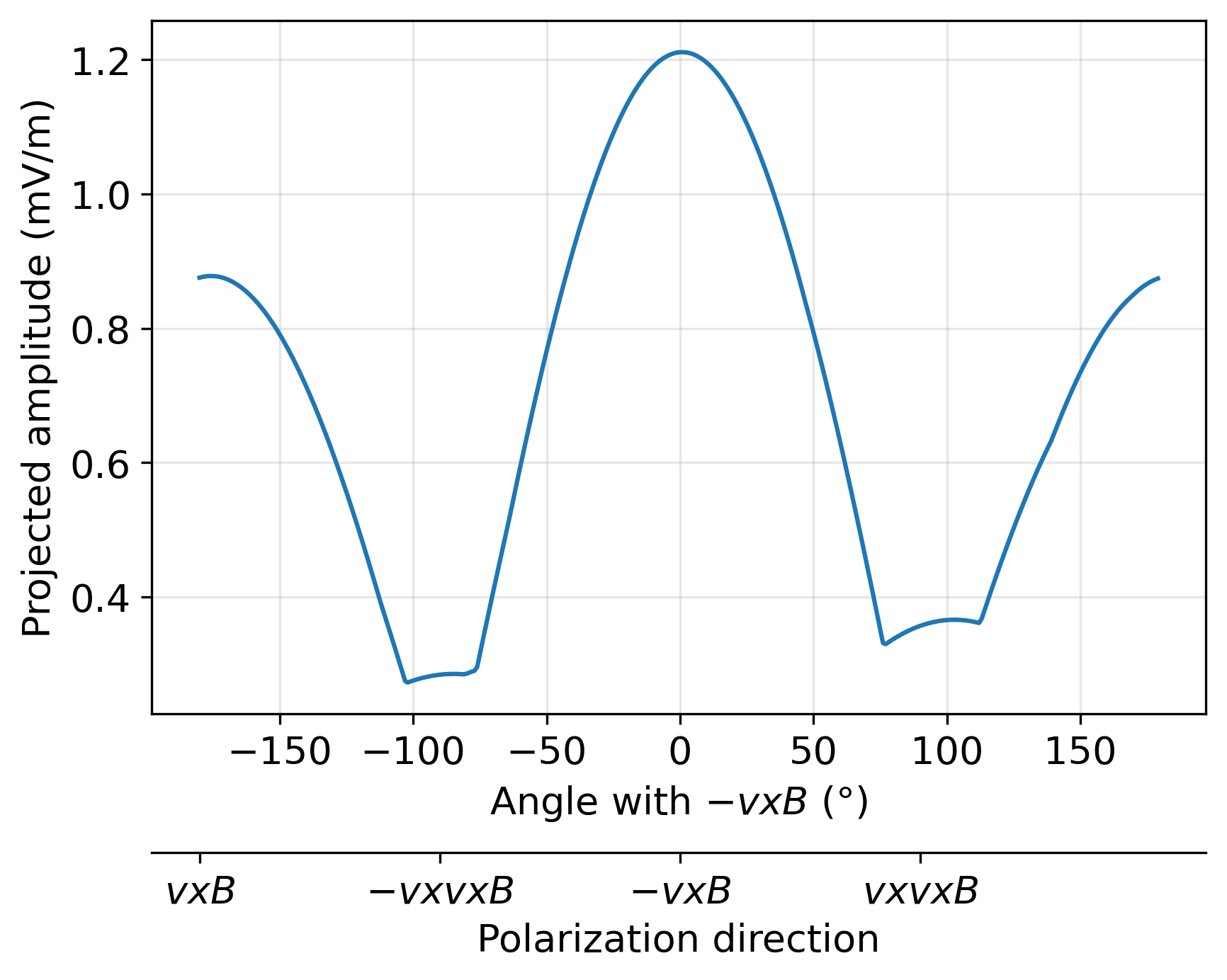}
    \includegraphics[width=0.49\linewidth]{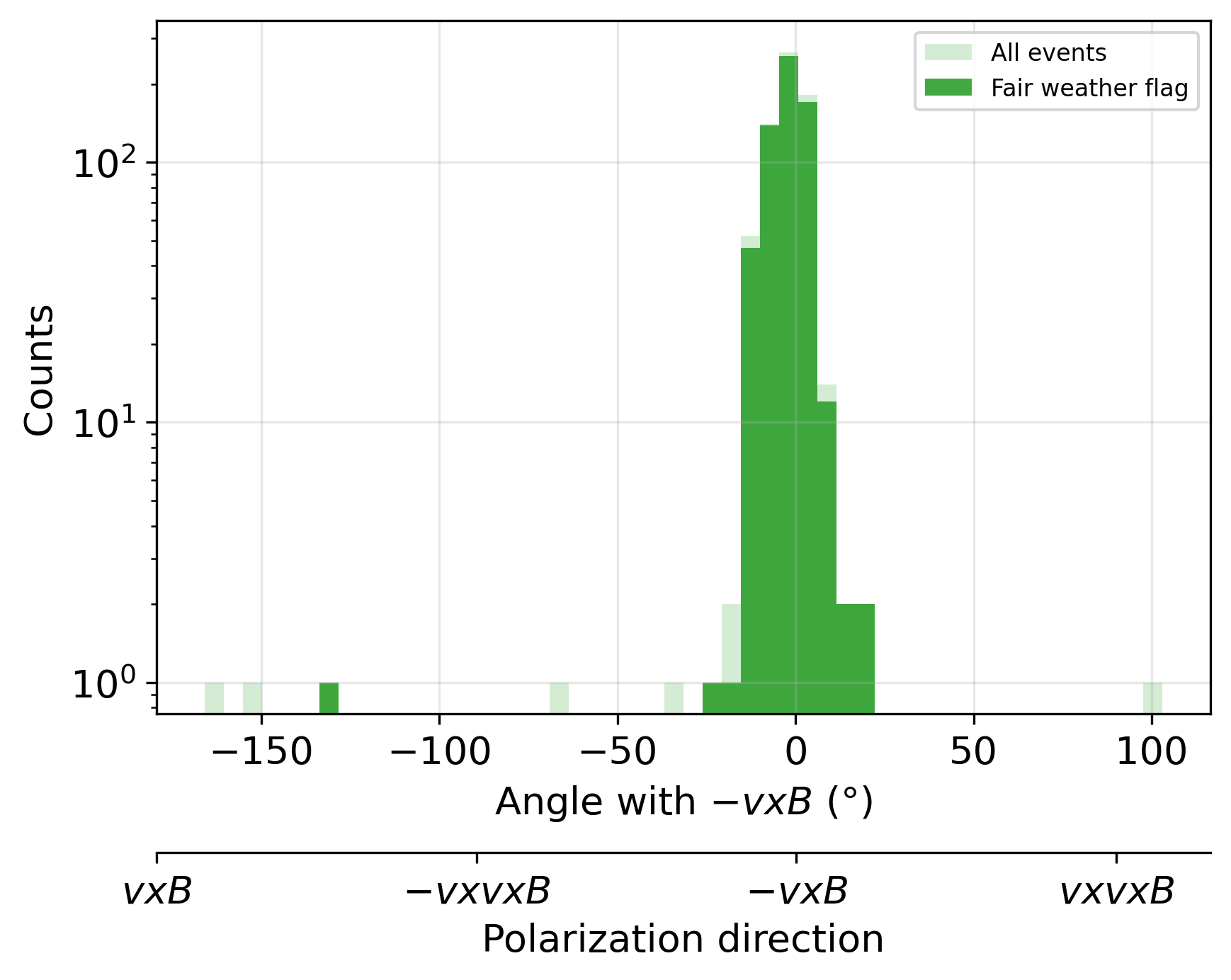}
    \caption{Left, example of projected amplitude of the summed electric field vector as a function of angle with respect to the $-\vec{v}\times\vec{B}$ direction.  Right: distributions of angles at the which the maximum angle appears. The mean and standard deviation of the distributions of \textit{All events} and \textit{Fair weather flag} are respectively -2.8\,$^\circ$ $\pm$ 12.2\,$^\circ$ and -2.1\,$^\circ$ $\pm$ 5.0\,$^\circ$. The one outlier at $-135^\circ$, was not flagged by the standard flag, however close inspection of weather data revealed thunderstorm conditions at that time.}
    \label{fig:polarisation}    
\end{figure}
Summing the waveforms of individual antennas provides a way to assess the polarisation of the radio signal on an event level, reducing the influence of noise when considering the polarization of the signal at each station. The radio emission mechanism arises mainly from the deflection of electrons and positrons in the geomagnetic field, resulting in an electric field polarised in the $-\vec{v}\times {B}$ direction, which we have been using in the previous section. For the selected events, at the location of $X_\text{RIT}$, we project the summed waveform onto a vector that we rotate in the plane perpendicular to the shower axis. An example of this amplitude as a function of the rotation angle is shown in Figure \ref{fig:polarisation} (left). The distribution of angles corresponding to the maximum amplitude for our selected events, including and excluding weather conditions for thunderstorms, is shown in Figure \ref{fig:polarisation} (right). As we can see, the preferred polarisation direction is centred around the expectations for the geomagnetic emission. We note that, unlike earlier studies in this frequency band~\cite{2014PhRvD..89e2002A}, the polarity of the signal is also determined. We find that in all fair-weather events, the emission is indeed polarised in the $-\vec{v}\times \vec{B}$ direction (and not in the $\vec{v}\times \vec{B}$ direction). These findings are particularly interesting from the point of view of the anomalous ANITA events~\cite{Anita}, where the observed polarity had an opposite sign compared to the expectation of reflected radio emission from the air shower off the Antarctic ice sheet. 
 
\section{Low-energy application}
Since with interferometry, the signal-to-noise ratio (when defined linear) scales with the square root of the number of antennas~\cite{schoorlemmer-2021}, it is possible to retrieve the air-shower signal even though it is not clearly present in the waveforms measured by individual antennas. Therefore, the detection threshold can be lowered, thereby lowering the energy threshold of the cosmic ray that can be detected. 
\begin{figure}
    \centering
    \includegraphics[width=0.49\linewidth]{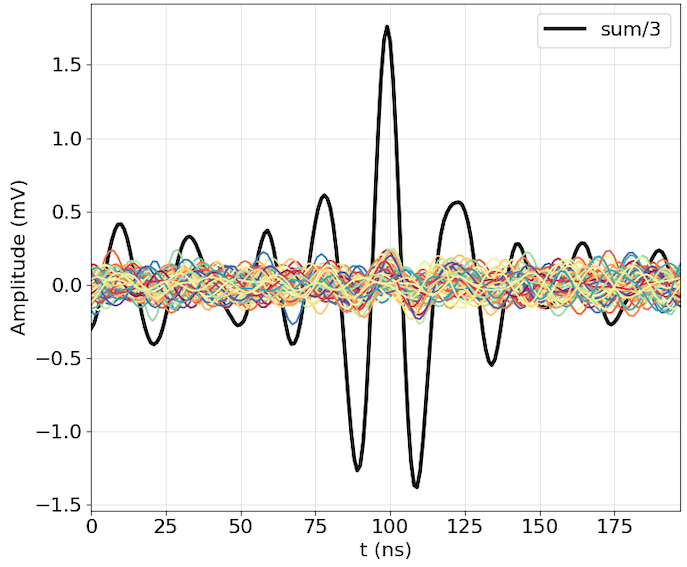}
    \includegraphics[width=0.49\linewidth]{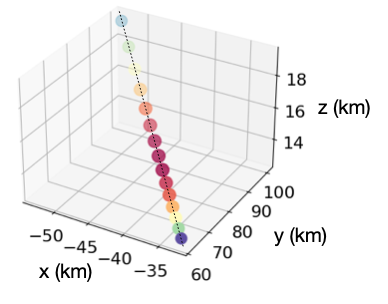}
        \caption{Example of interferometric reconstruction of an air shower that did not pass the standard \emph{Radio} reconstruction. Left: the individual waveforms (coloured lines) and the summed waveform (black line) at the location that resulted in maximum coherence. Right: axis reconstruction of this air shower using the interferometric technique, the marker colours indicate the relative intensity at each slice (red -high, blue-low).}
    \label{fig:subthreshold}
\end{figure}
Based solely on the SD-750 reconstruction, we selected a set of high-zenith angle ($75^\circ<\theta<82^\circ$) air showers, which are expected to illuminate more than 60 AERA radio antennas. From the events, we selected, based on the \emph{SD} energy estimator (see~\cite{SDReco}), air showers with an energy of $\sim$1\,EeV, which is near the threshold where we start to get efficiency for the \emph{Radio} reconstruction. This selection resulted in 194 air showers, on which the \emph{Radio} reconstruction was able to reconstruct 17 events close to the direction estimated with the SD-750 reconstruction, while the interferometric reconstruction of these events resulted in 57 events. Since we used the SD-750 reconstruction for starting values, we identified the reconstruction successful when a signal-to-noise ratio\footnote{Here SNR is defined as defined as the maximum amplitude divided by the standard deviation of noise window near the location of the maximum} larger than 7 was obtained in the summed waveform. Of the 17 radio-reconstructed events, two did not pass the interferometric reconstruction, manual inspection revealed that some high-noise stations might cause this failure. Of the successful interferometric reconstructed events, 68\% were within 0.63$^\circ$ (530\,m) from the SD-750 direction (core location) reconstruction. Here, we note that we were unable to correct for the known bias and that the event quality selection was not as strict as described in Section 2.3.   
An example of an event that did not have \emph{Radio} reconstruction of the air shower axis but was reconstructed with radio interferometry is shown in Figure \ref{fig:subthreshold}.

\section{Conclusion}
\begin{wrapfigure}{r}{0.5\textwidth}
  \vspace{-10pt}
  \begin{center}
    \includegraphics[trim={0 0.8cm 0.5cm 0.5cm},clip,width=0.5\textwidth]{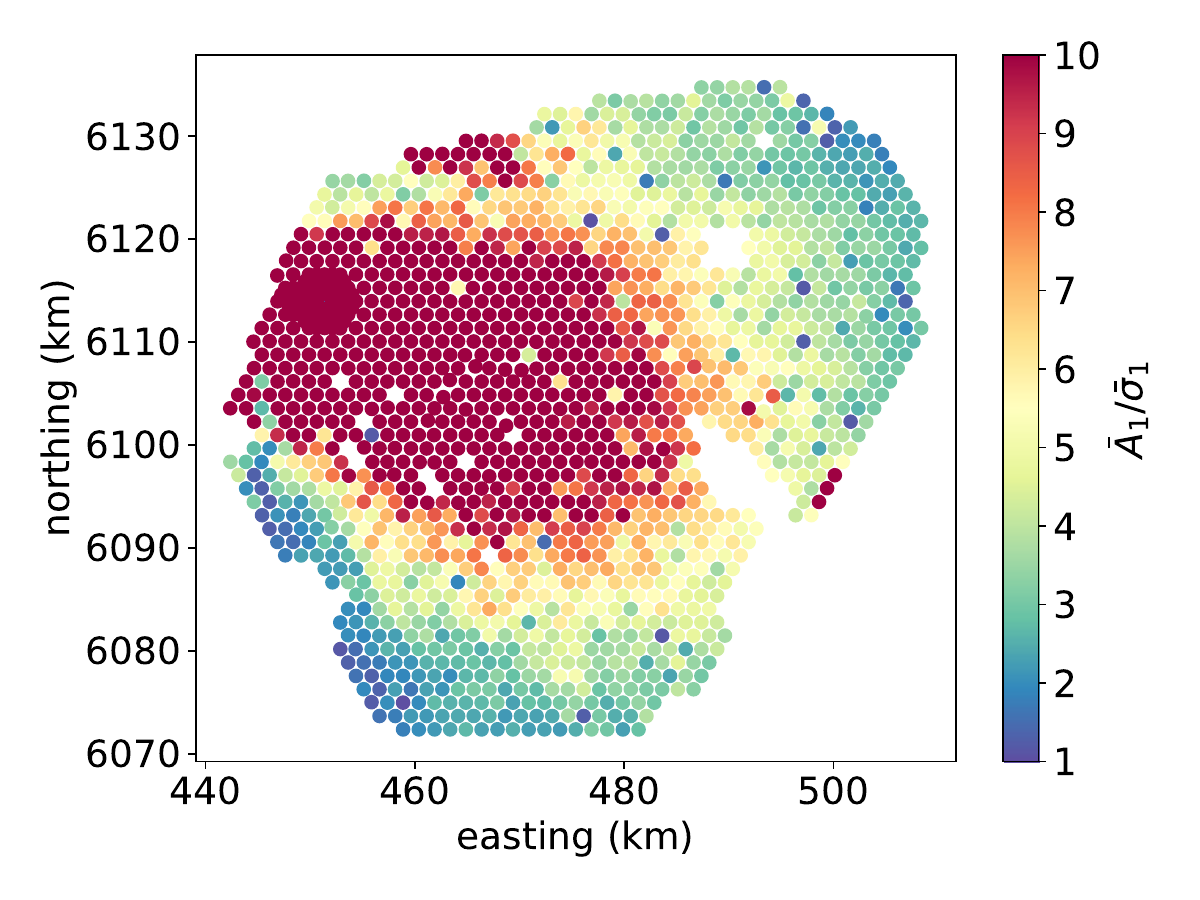}
  \end{center}  
  \caption{Measurements of a prototype beacon for the RD. Shown in colour is the Signal-to-Noise Ratio (mean amplitude $\bar{A}_1$ divided by mean noise level $\bar{\sigma}_1$). The colour range has been saturated at a value of 10.}
  \label{fig:beacon}
  \vspace{-12pt}
\end{wrapfigure}
For the first time, we presented the results of interferometric air-shower reconstruction on hundreds of events. The proposed method provides a robust and improved (or at least comparable) performance for geometry reconstruction compared to traditional methods. In addition, we showed a novel way of looking at the polarization of air showers, which confirms the $-\vec{v}\times\vec{B}$ polarization in inclined air showers.
In addition, we showed the potential of the method to reduce the energy threshold for air-shower reconstruction, where many antennas are illuminated. This resulted in roughly a    threefold more reconstructable events compared to the conventional radio reconstruction. This result is of particular interest for the application of the interferometric method for air-shower observations with the Square Kilometre Array~\cite{SKA} and the Southern Wide-view Gamma-ray Observatory~\cite{SWGO}.
These results were made possible with AERA due to its time synchronisation system and accurately measured antenna locations. With the recent (November 2024) finished upgrade of the Pierre Auger Observatory, \mbox{AugerPrime}, the surface detector units are extended with a Surface Scintillator Detector (SSD) and a Radio Detector (RD). We also plan and have already secured funding to expand the use of radio interferometry to the full array, enabling high-precision reconstruction of the highest energetic cosmic-ray air showers and getting a measure of the depth of shower maximum for inclined air showers. For this, the clocks of the radio data acquisition need to be synchronised, for which we plan to install a more powerful beacon system in the coming year. To validate its performance, a dedicated test system based on White Rabbit technology has been developed to assess the synchronisation accuracy of the beacon on the full array scale. A test measurement of a prototype beacon, which transmits a 50.4\,MHz line with a 50\,W amplifier, is shown in Figure \ref{fig:beacon}. In addition, we will carry out a differential GPS survey of the full array to localise the antennas to the required accuracy.

\clearpage


%
%
%

\section*{The Pierre Auger Collaboration}

{\footnotesize\setlength{\baselineskip}{10pt}
\noindent
\begin{wrapfigure}[11]{l}{0.12\linewidth}
\vspace{-4pt}
\includegraphics[width=0.98\linewidth]{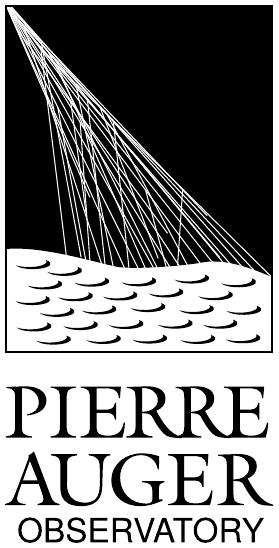}
\end{wrapfigure}
\begin{sloppypar}\noindent
A.~Abdul Halim$^{13}$,
P.~Abreu$^{70}$,
M.~Aglietta$^{53,51}$,
I.~Allekotte$^{1}$,
K.~Almeida Cheminant$^{78,77}$,
A.~Almela$^{7,12}$,
R.~Aloisio$^{44,45}$,
J.~Alvarez-Mu\~niz$^{76}$,
A.~Ambrosone$^{44}$,
J.~Ammerman Yebra$^{76}$
% ...

\end{sloppypar}
\begin{center}
\end{center}

\vspace{1ex}

\begin{description}[labelsep=0.2em,align=right,labelwidth=0.7em,labelindent=0em,leftmargin=2em,noitemsep,before={\renewcommand\makelabel[1]{##1 }}]
\item[$^{1}$] Centro At\'omico Bariloche and Instituto Balseiro (CNEA-UNCuyo-CONICET), San Carlos de Bariloche, Argentina
\item[$^{2}$] Departamento de F\'\i{}sica and Departamento de Ciencias de la Atm\'osfera y los Oc\'eanos, FCEyN, Universidad de Buenos Aires and CONICET, Buenos Aires, Argentina
\item[$^{3}$] IFLP, Universidad Nacional de La Plata and CONICET, La Plata, Argentina
\item[$^{4}$] Instituto de Astronom\'\i{}a y F\'\i{}sica del Espacio (IAFE, CONICET-UBA), Buenos Aires, Argentina
\item[$^{5}$] Instituto de F\'\i{}sica de Rosario (IFIR) -- CONICET/U.N.R.\ and Facultad de Ciencias Bioqu\'\i{}micas y Farmac\'euticas U.N.R., Rosario, Argentina
\item[$^{6}$] Instituto de Tecnolog\'\i{}as en Detecci\'on y Astropart\'\i{}culas (CNEA, CONICET, UNSAM), and Universidad Tecnol\'ogica Nacional -- Facultad Regional Mendoza (CONICET/CNEA), Mendoza, Argentina
\item[$^{7}$] Instituto de Tecnolog\'\i{}as en Detecci\'on y Astropart\'\i{}culas (CNEA, CONICET, UNSAM), Buenos Aires, Argentina
\item[$^{8}$] International Center of Advanced Studies and Instituto de Ciencias F\'\i{}sicas, ECyT-UNSAM and CONICET, Campus Miguelete -- San Mart\'\i{}n, Buenos Aires, Argentina
\item[$^{9}$] Laboratorio Atm\'osfera -- Departamento de Investigaciones en L\'aseres y sus Aplicaciones -- UNIDEF (CITEDEF-CONICET), Argentina
\item[$^{10}$] Observatorio Pierre Auger, Malarg\"ue, Argentina
\item[$^{11}$] Observatorio Pierre Auger and Comisi\'on Nacional de Energ\'\i{}a At\'omica, Malarg\"ue, Argentina
\item[$^{12}$] Universidad Tecnol\'ogica Nacional -- Facultad Regional Buenos Aires, Buenos Aires, Argentina
\item[$^{13}$] University of Adelaide, Adelaide, S.A., Australia
\item[$^{14}$] Universit\'e Libre de Bruxelles (ULB), Brussels, Belgium
\item[$^{15}$] Vrije Universiteit Brussels, Brussels, Belgium
\item[$^{16}$] Centro Brasileiro de Pesquisas Fisicas, Rio de Janeiro, RJ, Brazil
\item[$^{17}$] Centro Federal de Educa\c{c}\~ao Tecnol\'ogica Celso Suckow da Fonseca, Petropolis, Brazil
\item[$^{18}$] Instituto Federal de Educa\c{c}\~ao, Ci\^encia e Tecnologia do Rio de Janeiro (IFRJ), Brazil
\item[$^{19}$] Universidade de S\~ao Paulo, Escola de Engenharia de Lorena, Lorena, SP, Brazil
\item[$^{20}$] Universidade de S\~ao Paulo, Instituto de F\'\i{}sica de S\~ao Carlos, S\~ao Carlos, SP, Brazil
\item[$^{21}$] Universidade de S\~ao Paulo, Instituto de F\'\i{}sica, S\~ao Paulo, SP, Brazil
\item[$^{22}$] Universidade Estadual de Campinas (UNICAMP), IFGW, Campinas, SP, Brazil
\item[$^{23}$] Universidade Estadual de Feira de Santana, Feira de Santana, Brazil
\item[$^{24}$] Universidade Federal de Campina Grande, Centro de Ciencias e Tecnologia, Campina Grande, Brazil
\item[$^{25}$] Universidade Federal do ABC, Santo Andr\'e, SP, Brazil
\item[$^{26}$] Universidade Federal do Paran\'a, Setor Palotina, Palotina, Brazil
\item[$^{27}$] Universidade Federal do Rio de Janeiro, Instituto de F\'\i{}sica, Rio de Janeiro, RJ, Brazil
\item[$^{28}$] Universidad de Medell\'\i{}n, Medell\'\i{}n, Colombia
\item[$^{29}$] Universidad Industrial de Santander, Bucaramanga, Colombia
\item[$^{30}$] Charles University, Faculty of Mathematics and Physics, Institute of Particle and Nuclear Physics, Prague, Czech Republic
\item[$^{31}$] Institute of Physics of the Czech Academy of Sciences, Prague, Czech Republic
\item[$^{32}$] Palacky University, Olomouc, Czech Republic
\item[$^{33}$] CNRS/IN2P3, IJCLab, Universit\'e Paris-Saclay, Orsay, France
\item[$^{34}$] Laboratoire de Physique Nucl\'eaire et de Hautes Energies (LPNHE), Sorbonne Universit\'e, Universit\'e de Paris, CNRS-IN2P3, Paris, France
\item[$^{35}$] Univ.\ Grenoble Alpes, CNRS, Grenoble Institute of Engineering Univ.\ Grenoble Alpes, LPSC-IN2P3, 38000 Grenoble, France
\item[$^{36}$] Universit\'e Paris-Saclay, CNRS/IN2P3, IJCLab, Orsay, France
\item[$^{37}$] Bergische Universit\"at Wuppertal, Department of Physics, Wuppertal, Germany
\item[$^{38}$] Karlsruhe Institute of Technology (KIT), Institute for Experimental Particle Physics, Karlsruhe, Germany
\item[$^{39}$] Karlsruhe Institute of Technology (KIT), Institut f\"ur Prozessdatenverarbeitung und Elektronik, Karlsruhe, Germany
\item[$^{40}$] Karlsruhe Institute of Technology (KIT), Institute for Astroparticle Physics, Karlsruhe, Germany
\item[$^{41}$] RWTH Aachen University, III.\ Physikalisches Institut A, Aachen, Germany
\item[$^{42}$] Universit\"at Hamburg, II.\ Institut f\"ur Theoretische Physik, Hamburg, Germany
\item[$^{43}$] Universit\"at Siegen, Department Physik -- Experimentelle Teilchenphysik, Siegen, Germany
\item[$^{44}$] Gran Sasso Science Institute, L'Aquila, Italy
\item[$^{45}$] INFN Laboratori Nazionali del Gran Sasso, Assergi (L'Aquila), Italy
\item[$^{46}$] INFN, Sezione di Catania, Catania, Italy
\item[$^{47}$] INFN, Sezione di Lecce, Lecce, Italy
\item[$^{48}$] INFN, Sezione di Milano, Milano, Italy
\item[$^{49}$] INFN, Sezione di Napoli, Napoli, Italy
\item[$^{50}$] INFN, Sezione di Roma ``Tor Vergata'', Roma, Italy
\item[$^{51}$] INFN, Sezione di Torino, Torino, Italy
\item[$^{52}$] Istituto di Astrofisica Spaziale e Fisica Cosmica di Palermo (INAF), Palermo, Italy
\item[$^{53}$] Osservatorio Astrofisico di Torino (INAF), Torino, Italy
\item[$^{54}$] Politecnico di Milano, Dipartimento di Scienze e Tecnologie Aerospaziali , Milano, Italy
\item[$^{55}$] Universit\`a del Salento, Dipartimento di Matematica e Fisica ``E.\ De Giorgi'', Lecce, Italy
\item[$^{56}$] Universit\`a dell'Aquila, Dipartimento di Scienze Fisiche e Chimiche, L'Aquila, Italy
\item[$^{57}$] Universit\`a di Catania, Dipartimento di Fisica e Astronomia ``Ettore Majorana``, Catania, Italy
\item[$^{58}$] Universit\`a di Milano, Dipartimento di Fisica, Milano, Italy
\item[$^{59}$] Universit\`a di Napoli ``Federico II'', Dipartimento di Fisica ``Ettore Pancini'', Napoli, Italy
\item[$^{60}$] Universit\`a di Palermo, Dipartimento di Fisica e Chimica ''E.\ Segr\`e'', Palermo, Italy
\item[$^{61}$] Universit\`a di Roma ``Tor Vergata'', Dipartimento di Fisica, Roma, Italy
\item[$^{62}$] Universit\`a Torino, Dipartimento di Fisica, Torino, Italy
\item[$^{63}$] Benem\'erita Universidad Aut\'onoma de Puebla, Puebla, M\'exico
\item[$^{64}$] Unidad Profesional Interdisciplinaria en Ingenier\'\i{}a y Tecnolog\'\i{}as Avanzadas del Instituto Polit\'ecnico Nacional (UPIITA-IPN), M\'exico, D.F., M\'exico
\item[$^{65}$] Universidad Aut\'onoma de Chiapas, Tuxtla Guti\'errez, Chiapas, M\'exico
\item[$^{66}$] Universidad Michoacana de San Nicol\'as de Hidalgo, Morelia, Michoac\'an, M\'exico
\item[$^{67}$] Universidad Nacional Aut\'onoma de M\'exico, M\'exico, D.F., M\'exico
\item[$^{68}$] Institute of Nuclear Physics PAN, Krakow, Poland
\item[$^{69}$] University of \L{}\'od\'z, Faculty of High-Energy Astrophysics,\L{}\'od\'z, Poland
\item[$^{70}$] Laborat\'orio de Instrumenta\c{c}\~ao e F\'\i{}sica Experimental de Part\'\i{}culas -- LIP and Instituto Superior T\'ecnico -- IST, Universidade de Lisboa -- UL, Lisboa, Portugal
\item[$^{71}$] ``Horia Hulubei'' National Institute for Physics and Nuclear Engineering, Bucharest-Magurele, Romania
\item[$^{72}$] Institute of Space Science, Bucharest-Magurele, Romania
\item[$^{73}$] Center for Astrophysics and Cosmology (CAC), University of Nova Gorica, Nova Gorica, Slovenia
\item[$^{74}$] Experimental Particle Physics Department, J.\ Stefan Institute, Ljubljana, Slovenia
\item[$^{75}$] Universidad de Granada and C.A.F.P.E., Granada, Spain
\item[$^{76}$] Instituto Galego de F\'\i{}sica de Altas Enerx\'\i{}as (IGFAE), Universidade de Santiago de Compostela, Santiago de Compostela, Spain
\item[$^{77}$] IMAPP, Radboud University Nijmegen, Nijmegen, The Netherlands
\item[$^{78}$] Nationaal Instituut voor Kernfysica en Hoge Energie Fysica (NIKHEF), Science Park, Amsterdam, The Netherlands
\item[$^{79}$] Stichting Astronomisch Onderzoek in Nederland (ASTRON), Dwingeloo, The Netherlands
\item[$^{80}$] Universiteit van Amsterdam, Faculty of Science, Amsterdam, The Netherlands
\item[$^{81}$] Case Western Reserve University, Cleveland, OH, USA
\item[$^{82}$] Colorado School of Mines, Golden, CO, USA
\item[$^{83}$] Department of Physics and Astronomy, Lehman College, City University of New York, Bronx, NY, USA
\item[$^{84}$] Michigan Technological University, Houghton, MI, USA
\item[$^{85}$] New York University, New York, NY, USA
\item[$^{86}$] University of Chicago, Enrico Fermi Institute, Chicago, IL, USA
\item[$^{87}$] University of Delaware, Department of Physics and Astronomy, Bartol Research Institute, Newark, DE, USA
\item[] -----
\item[$^{a}$] Max-Planck-Institut f\"ur Radioastronomie, Bonn, Germany
\item[$^{b}$] also at Kapteyn Institute, University of Groningen, Groningen, The Netherlands
\item[$^{c}$] School of Physics and Astronomy, University of Leeds, Leeds, United Kingdom
\item[$^{d}$] Fermi National Accelerator Laboratory, Fermilab, Batavia, IL, USA
\item[$^{e}$] Pennsylvania State University, University Park, PA, USA
\item[$^{f}$] Colorado State University, Fort Collins, CO, USA
\item[$^{g}$] Louisiana State University, Baton Rouge, LA, USA
\item[$^{h}$] now at Graduate School of Science, Osaka Metropolitan University, Osaka, Japan
\item[$^{i}$] Institut universitaire de France (IUF), France
\item[$^{j}$] now at Technische Universit\"at Dortmund and Ruhr-Universit\"at Bochum, Dortmund and Bochum, Germany
\end{description}

\section*{Acknowledgments}

\begin{sloppypar}
The successful installation, commissioning, and operation of the Pierre
Auger Observatory would not have been possible without the strong
commitment and effort from the technical and administrative staff in
Malarg\"ue. We are very grateful to the following agencies and
organizations for financial support:
\end{sloppypar}

\begin{sloppypar}
Argentina -- Comisi\'on Nacional de Energ\'\i{}a At\'omica; Agencia Nacional de
Promoci\'on Cient\'\i{}fica y Tecnol\'ogica (ANPCyT); Consejo Nacional de
Investigaciones Cient\'\i{}ficas y T\'ecnicas (CONICET); Gobierno de la
Provincia de Mendoza; Municipalidad de Malarg\"ue; NDM Holdings and Valle
Las Le\~nas; in gratitude for their continuing cooperation over land
access; Australia -- the Australian Research Council; Belgium -- Fonds
de la Recherche Scientifique (FNRS); Research Foundation Flanders (FWO),
Marie Curie Action of the European Union Grant No.~101107047; Brazil --
Conselho Nacional de Desenvolvimento Cient\'\i{}fico e Tecnol\'ogico (CNPq);
Financiadora de Estudos e Projetos (FINEP); Funda\c{c}\~ao de Amparo \`a
Pesquisa do Estado de Rio de Janeiro (FAPERJ); S\~ao Paulo Research
Foundation (FAPESP) Grants No.~2019/10151-2, No.~2010/07359-6 and
No.~1999/05404-3; Minist\'erio da Ci\^encia, Tecnologia, Inova\c{c}\~oes e
Comunica\c{c}\~oes (MCTIC); Czech Republic -- GACR 24-13049S, CAS LQ100102401,
MEYS LM2023032, CZ.02.1.01/0.0/0.0/16{\textunderscore}013/0001402,
CZ.02.1.01/0.0/0.0/18{\textunderscore}046/0016010 and
CZ.02.1.01/0.0/0.0/17{\textunderscore}049/0008422 and CZ.02.01.01/00/22{\textunderscore}008/0004632;
France -- Centre de Calcul IN2P3/CNRS; Centre National de la Recherche
Scientifique (CNRS); Conseil R\'egional Ile-de-France; D\'epartement
Physique Nucl\'eaire et Corpusculaire (PNC-IN2P3/CNRS); D\'epartement
Sciences de l'Univers (SDU-INSU/CNRS); Institut Lagrange de Paris (ILP)
Grant No.~LABEX ANR-10-LABX-63 within the Investissements d'Avenir
Programme Grant No.~ANR-11-IDEX-0004-02; Germany -- Bundesministerium
f\"ur Bildung und Forschung (BMBF); Deutsche Forschungsgemeinschaft (DFG);
Finanzministerium Baden-W\"urttemberg; Helmholtz Alliance for
Astroparticle Physics (HAP); Helmholtz-Gemeinschaft Deutscher
Forschungszentren (HGF); Ministerium f\"ur Kultur und Wissenschaft des
Landes Nordrhein-Westfalen; Ministerium f\"ur Wissenschaft, Forschung und
Kunst des Landes Baden-W\"urttemberg; Italy -- Istituto Nazionale di
Fisica Nucleare (INFN); Istituto Nazionale di Astrofisica (INAF);
Ministero dell'Universit\`a e della Ricerca (MUR); CETEMPS Center of
Excellence; Ministero degli Affari Esteri (MAE), ICSC Centro Nazionale
di Ricerca in High Performance Computing, Big Data and Quantum
Computing, funded by European Union NextGenerationEU, reference code
CN{\textunderscore}00000013; M\'exico -- Consejo Nacional de Ciencia y Tecnolog\'\i{}a
(CONACYT) No.~167733; Universidad Nacional Aut\'onoma de M\'exico (UNAM);
PAPIIT DGAPA-UNAM; The Netherlands -- Ministry of Education, Culture and
Science; Netherlands Organisation for Scientific Research (NWO); Dutch
national e-infrastructure with the support of SURF Cooperative; Poland
-- Ministry of Education and Science, grants No.~DIR/WK/2018/11 and
2022/WK/12; National Science Centre, grants No.~2016/22/M/ST9/00198,
2016/23/B/ST9/01635, 2020/39/B/ST9/01398, and 2022/45/B/ST9/02163;
Portugal -- Portuguese national funds and FEDER funds within Programa
Operacional Factores de Competitividade through Funda\c{c}\~ao para a Ci\^encia
e a Tecnologia (COMPETE); Romania -- Ministry of Research, Innovation
and Digitization, CNCS-UEFISCDI, contract no.~30N/2023 under Romanian
National Core Program LAPLAS VII, grant no.~PN 23 21 01 02 and project
number PN-III-P1-1.1-TE-2021-0924/TE57/2022, within PNCDI III; Slovenia
-- Slovenian Research Agency, grants P1-0031, P1-0385, I0-0033, N1-0111;
Spain -- Ministerio de Ciencia e Innovaci\'on/Agencia Estatal de
Investigaci\'on (PID2019-105544GB-I00, PID2022-140510NB-I00 and
RYC2019-027017-I), Xunta de Galicia (CIGUS Network of Research Centers,
Consolidaci\'on 2021 GRC GI-2033, ED431C-2021/22 and ED431F-2022/15),
Junta de Andaluc\'\i{}a (SOMM17/6104/UGR and P18-FR-4314), and the European
Union (Marie Sklodowska-Curie 101065027 and ERDF); USA -- Department of
Energy, Contracts No.~DE-AC02-07CH11359, No.~DE-FR02-04ER41300,
No.~DE-FG02-99ER41107 and No.~DE-SC0011689; National Science Foundation,
Grant No.~0450696, and NSF-2013199; The Grainger Foundation; Marie
Curie-IRSES/EPLANET; European Particle Physics Latin American Network;
and UNESCO.
\end{sloppypar}

}

\end{document}